\documentclass[preprintnumbers,amsmath,12pt,amssymb,floatfix,superscriptaddress,nofootinbib]{article}

\def\mytitle#1{\setcounter{equation}{0}
\setcounter{footnote}{0}
\begin{center}\Large\textbf{#1}\end{center}
\vspace{0.25cm}}
\def\myname#1{\begin{center}{\large #1}\end{center}\vspace{-0.13cm}}
\def\myplace#1#2{\small\begin{flushleft}\textit{#1}\\
\texttt{#2}\end{flushleft}}

\def\myclassification#1{\small\noindent
Keywords :  #1\vspace{0.5cm}}
\usepackage{amsmath}
\usepackage[table]{xcolor}
\usepackage{color}
\usepackage{amsfonts}
\usepackage{amssymb}
\usepackage{breqn}
\usepackage{array}
\usepackage{graphicx}
\usepackage[utf8]{inputenc}
\usepackage{hyperref}

 \date{}
 
\begin{document}
\mytitle{Einstein-Maxwell-Scalar Black Hole: Thermodynamic Properties with Logarithmically Corrected Barrow's Fractalised Entropy}

\myname{Ritabrata~Biswas*\footnote{biswas.ritabrata@gmail.com~~;~~\text{Orchid}~:~0000-0003-3086-892X} and Satyajit~Pal*,**\footnote{satyajitpal1995@gmail.com~~;~~\text{Orchid}~:~0009-0009-2099-4501}}
%\myname{$\text{Ritabrata~Biswas}^*$ \footnote{biswas.ritabrata@gmail.com~~;~~\text{Orchid}~:~0000-0003-3086-892X} and $\text{Satyajit ~Pal}^{*,**}$\footnote{satyajitpal1995@gmail.com~~;~~\text{Orchid}~:~0009-0009-2099-4501}
\vspace{0.5cm}
\myplace{*Department of Mathematics, The University of Burdwan, Burdwan-713104, India\\ **Department of Mathematics, Dr. Bhupendra Nath Dutta Smriti Mahavidyalaya, Hatgobindapur, Purba Bardhaman-713407, India}{}

%%%%%%%%%%%%%%%%%%%%%%%%%%%%%%%%%%%%%%%%%%%%%%%%%%%%%%%%%%%%%%%%%%%%
\begin{abstract}
The thermodynamics of black holes (BHs) within the Einstein–Maxwell–Scalar (EMS) framework, incorporating Barrow's fractalised entropy($S_B$) and its logarithmic correction ($S_{BLC}$) to analyze quantum gravity effects are investigated in this article. A static, spherically symmetric BH solution is obtained by coupling the scalar field nonminimally to the electromagnetic field through a scalar-dependent function. The thermodynamic properties $-$ including temperature, specific heat and Gibbs free energy $-$ are derived and explored in the context of Barrow-modified entropy. Phase transitions and critical behavior are followed by analyzing $P–V$ criticality and the influence of scalar and electric charges on stability. Furthermore, the Joule–Thomson(JT) expansion is examined to understand the inversion behavior and thermodynamic responses under adiabatic expansion. Geometrothermodynamic Ricci is also studied. Our findings suggest the presence of thermodynamic instabilities, remnants and nontrivial critical phenomena, providing new insights into BH thermodynamics, information loss paradox, a probable explanation towards the formation of dark energy and dark matter etc in modified gravity scenario with quantum corrections.
\end{abstract}
%%%%%%%%%%%%%%%%%%%%%%%%%%%%%%%%%%%%%%%%%%%%%%%%%%%%%%%%%%%%%%%%%%%%
\myclassification{Einstein–Maxwell–Scalar theory, Fractalised entropy of black holes, Logarithmic corrections to entropy, Phase transitions in Black Hole thermodynamics, $P–V$ criticality in Black Hole thermodynamics, Joule–Thomson expansion in Black Hole thermodynamics}
%%%%%%%%%%%%%%%%%%%%%%%%%%%%%%%%%%%%%%%%%%%%%%%%%%%%%%%%%%%%%%%%%%%%%%%%%%%%%%%%%%%
\section{Introduction}
%%%%%%%%%%%%%%%%%%%%%%%%%%%%%%%%%%%%%%%%%%%%%%%%%%%%%%%%%%%%%%%%%%
A black hole (BH) is such a dense object that its gravity beneath its surface, popularly coined as the event horizon, turns strong enough to block electromagnetic waves to escape. However, energy released by nearby objects which are loosing their gravitational potential due to the fall towards the central gravitating core can be observed in different wavelengths.

In 2019, Event Horizon Telescope measured a $6.5\pm 0.7$ billion solar masses at the center of Meisser 87. This giant was residing only in approximately 40 billion kilometers. This object creates significant shadow and is nothing but a BH \cite{ medeiros2023image}. In 2020, a 20 microsecond resolution image of the blazar $3C279$ was explained \cite{gomez2020event} which was nothing but a giant BH. Center of the Centaurus A was imaged in 2021 (Analysis of motion of objects are studied in the nearby region centering where we expect the presence of a massive BH). They are direct evidences of existences of BHs\cite{ janssen2021event}.

Alongside different tensor approaches to construct an appropriate gravity theory, inclusion of scalar field is found. Fisher's study \cite{fisher1999scalar} of massless scalar field solution was the first one in this row. Incorporation of dilaton and axion fields to the Einstein's action is observed to get the effects of supergravity and string theory \cite{schwarz1987witten,beissen2024geometrothermodynamics,harms1992statistical}.

Anti-deSitter (AdS) space  - Conformal Field Theory correspondence opens the path to find an asymptotically AdS BH solution in dilaton gravity \cite{witten1998anti,singh2024cosmic,hendi2010thermodynamics,davies1978thermodynamics, middelburgthermodynamics}. Various self interaction potentials, dilaton potential, Cosmological Constant  ($\Lambda$) etc are found and related BHs are tested for phase transitions and holographic thermal and thermodynamic phenomena \cite{sheykhi2010thermodynamic, yasir2021thermodynamic,mustafa2022shadows,christodoulou1999instability}.

BHs are followed to possess some properties analogous to classical thermodynamic entities which follow laws analogous to the laws of classical thermodynamics. Pioneers of this study were Jacob Bekenstein, Brandon Carter, James Bardeen, Stephen Hawking etc \cite{bekenstein2020black, hawking1975particle}. 

 Using the old entropy formula, quantum gravitational corrections to a BH's entropy is done in the article \cite{calmet2021quantum} where the correction term is interpreted as a pressure term in the first law of BH thermodynamics. In this work, for lightest BH of Planckian masses, the thermodynamic properties are found much more  classical than naively expected. Another article \cite{sekhmani2025extended} studied the extended phase space thermodynamics where critical behavior of charged AdS BHs surrounded by polytropic scalar gas field are analyzed.  Series of geothermodynamic tools, e.g Weinhole, Ruppinier, Quevedo I, II, ``Hendi, Panahiyan, Eslam Panah, and Momennia"(HDEM)  are utilized in this article. Next,  we  will discuss regarding the article \cite{bouzenada2025barrow} where the authors have examined different test particles' behavior around a non rotating Frolov BH. Barrow entropy effect is analyzed on different thermodynamic parameters. The Schwarzschild geometry, in the context of effective field theory models of gravity, is studied in the article \cite{battista2024quantum}. Here, quantum matrices are followed not to satisfy the condition  $ - g_{tt} = g^{rr}$. The analysis shows that there is no choice of the sign of the constant parameter embodying the quantum correction to a metric which leaves all the features of the classical Schwarzschild domain almost unaffected. 

Thermodynamics and phase transition of asymptotically AdS BHs in the presence of ModMax nonlinear electrodynamics and perfect fluid dark matter. Geometrothermodynamics are studied including the Weinhole, Ruppiner, HDEM metrices \cite{sekhmani2025thermodynamics}. First order quantum correction of thermodynamics in a charged accelerating AdS BH with gauge potential is carried in \cite{ali2024first}. Examination of the spherical trajectory of test particles surrounding a nonrotating BH within the context of Symmergent gravity is studied in the article\cite{maurya2025circular}. Dilaton BHs with logarithmic source in gravity's rainbow is studied in the article \cite{dehghani2021dilaton}. Again quasinormal modes and greybody bounds of BHs embedded with modified Chaplygin gas type dark energy is found in literature \cite{sekhmani2025quasinormal}. Strong gravitational lensing effects of various supermassive compact objects for the static and spherically symmetric hairy BH by gravitational decoupling is tested in the article  \cite{mustafa2024testing}. Authors of \cite{wang2025dynamical} have studied the shadow features of quantum Schwarzschild BH in effective theories of gravity.

With geometrized units, zeroth law is followed by a stationary BH's surface at its horizon which is found to show a nonnegative surface gravity $\kappa$. A small perturbation in mass/ energy $E$ of a BH follows
\begin{equation}\label{1}
\delta E=\frac{\kappa}{8\pi}\delta A_g+\Omega \delta J+\phi \delta Q~~,    
\end{equation}
where $E$, $A_g$, $\Omega$, $J$, $\phi$ and $Q$ are the energy, horizon area, angular velocity, angular momentum, electrostatic potential and electric charge respectively. Equation (\ref{1}) makes after the first law of thermodynamics. Second law of thermodynamics comes along with the weak energy condition stating the chronologically non-decreasing nature of the horizon area, i.e., $\frac{dA}{dt}\geq 0$. Third law of thermodynamics is followed from the fact that a vanishing surface gravity of a BH horizon is not at all probable.

In the next section(i.e. in section $\ref{BMEMS}$), we will discuss regarding the BH solution in Einstein Maxwell Scalar(EMS) framework and logarithmic correction to the Barrow entropy. In section $\ref{thermo}$, we will investigate the temperature, specific heat, free energy etc. Next, $P$-$V$ criticality will be analyzed in section $\ref{pv}$. After that, in section $\ref{jt}$, Joule-Thomson (JT) extension is studied. Section $\ref{GT}$ will briefly discuss regarding the geometrothermodynamics. Finally, brief discussion and conclusion is incorporated in the section $\ref{conclusion}$.

%%%%%%%%%%%%%%%%%%%%%%%%%%%%%%%%%%%%%%%%%%%%%%%%%%%%%%%%%%%%%%%%%%%%%%%%%%%%%%
\section{EMS BH and Barrow Entropy with Logarithmic Correction}\label{BMEMS}
%%%%%%%%%%%%%%%%%%%%%%%%%%%%%%%%%%%%%%%%%%%%%%%%%%%%%%%%%%%%%%%%%%%%%%%%%%%%%%
In Reissner Nordstrom AdS case, a BH is found to incorporate a trivial scalar field. In EMS theory, however, nonminimal coupling with the Maxwell invariant through $K(\phi)$ is considered. Scalar hair is also coupled. Phase transition beyond classical no hair theorem is considered as well \cite{zhang2025extraction,razina2019cosmological,wang2023observational}. Gravitational coupling to the electric charge $Q$ and scalar charge $D=-\frac{Q^2}{2M}$ \cite{astefanesei2019einstein} is observed.

The action of EMS theory is given by \cite{gao2004dilaton}
\begin{equation}\label{P4_EMS}
    S = \int d^4x \sqrt{-g} \left[ \mathcal{R} - 2\Delta_{\mu\varphi} \Delta^\mu _\varphi - K(\varphi)F^2 - V(\varphi) \right]~~~~,
\end{equation}
where $K(\varphi)$ denotes the Maxwell field, $F^2 = F_{\mu\nu}F^{\mu\nu}$, $F^{\mu\nu}$ is the electromagnetic field and ${\cal R}$ is the Ricci scalar curvature and $\varphi$ indicates the coupling function of the scalar field. Also $V(\varphi)$ represents the scalar potential. The representation of the BH space-time in a static, spherically symmetric form is given by
\begin{equation}
    ds^2 = - u(r)dt^2 + u(r)^{-1}dr^2 + f(r)(d\theta^2 + sin^2\theta d\varphi^2)~~~~.
\end{equation}
According to the theory, the mass of the dilaton BH in the deSitter universe \cite{gao2004dilaton} is $M$ when $K = e^{2\varphi}$. Thus,
\begin{equation}
    u(r) = 1- \frac{2M}{r}- \frac{1}{3}\lambda f,~~ f(r)= r\left( r - \frac{q^2}{M} \right),~~ \varphi = - \frac{1}{2}ln\left( 1-\frac{q^2}{Mr} \right)~~~,
\end{equation}
i.e.,
\begin{equation}\label{P4_lapse}
    u(r) = 1 -\frac{1}{3} \lambda  r
   \left(r-\frac{q^2}{M}\right)-\frac{2 M}{r}~~~~.
\end{equation}
The constant $\lambda$ plays a crucial role here. When $q = 0$, it goes back to the Schwarzschild-deSitter solution. To come up with a new solution, we examine how $u$ is formulated and find that the $\lambda$ term has a proportional relationship with $f$ rather than $r^2$. 

The Bekenstein Hawking entropy is \begin{equation}\label{entropy}
S=\frac{\kappa_B c^3 A_g}{4G_N\hbar}~~~~\text{and}
\end{equation} 
temperature is 
\begin{equation}\label{temp}
T= \frac{\hbar c^3}{8\pi G_N M \kappa_B}~~~~,
\end{equation}
where the units of the parameters are $\kappa_B(JK^{-1})$, $c^3(m^3s^{-3})$, $A_g(m^2)$, $G_N(m^3kg^{-1}s^{-2})$. Hence the entropy has the unit $JK^{-1}$. $G_N$ and $M$ are Newton's gravitational constant and the mass of the BH respectively. In natural units, however, entropy is a dimensionless pure number.

One of the primitive ideas regarding fractals, popularly known as the ``Koch Snowflake", was proposed in 1904 \cite{von2019continuous}. This intended to construct a two dimensional object constructed by an iterative process, to add self alike small parts which lead to a finite area and infinite perimeter. Sierpinski\cite{sierpinski1915curve} and Sponge \cite{menger2019general} have generalized this fractalization to three dimensional objects with finite volume and infinite surface area. John Barrow \cite{barrow2020area} proposed a similar fractal structure on static, spherically symmetric Schwarzschild BH caused by some quantum gravitational effects. $N$ number of hierarchically small ($\frac{r_{newer}}{r_{older}}=\tilde{\lambda}<1$) spheres are placed to touch the surface of the original sphere. After infinite steps, the volume and area will turn $V_{\infty}=\frac{4\pi}{3}R_g^3\sum_{n=0}^{\infty}\left(N\tilde{\lambda}^3\right)^n~~,~~A_{g\infty}=4\pi R_g^2\sum_{n=0}^{\infty}\left(N\tilde{\lambda}^2\right)^n$, where $R_g=2G_N M$, the Schwarzschild radius with speed of light as unit. $\tilde{\lambda}^{-2}<N<\tilde{\lambda}^{-2}$ is the required condition for volume to converge and area to diverge as $N\rightarrow \infty$. This Barrow statistics results the BH entropy as \cite{barrow2020area} 
\begin{equation}
S_{B}=\left(\frac{A_g}{A_{Pl}}\right)^{1+\frac{\Delta}{2}}~~~~,  
\end{equation}
the subscript $B$ denotes the Barrow entropy.

Surface area of Schwarzschild BH is denoted as $A_g=4\pi R_g^2$, $R_g$ being the radius of Schwarzschild Bh's event horizon. Planck area $A_{Pl}\sim 4G_N$. The condition $0\leq \Delta \leq 1$ reflects quantum gravity effects without containing any quantum parameter in it. Incorporating the surface area of the Schwarzschild BH, Barrow entropy is followed as
\begin{equation}
S_B=\left(4\pi G_N\right)^{1+\frac{\Delta}{2}}M^{2+\Delta}~~~~.    
\end{equation}
Considering self gravity and back reaction effects, a logarithmic correction \cite{banerjee2008quantum} can be done as 
\begin{equation}
S_{LC}=\left(\frac{A_g}{A_{Pl}}\right)+\alpha \ln \left(\frac{A_g}{A_{Pl}}\right)+\beta   ~~~~. 
\end{equation}
It is followed that the BH evaporation process takes a longer time that its counterpart in the logarithmic-corrected BH statistics. $\alpha$ and $\beta$ are constants.

Analysing the Schwarzschild BH, impacts of quantum gravity effects are considered by adding a logarithmic corrected term into Barrow entropy. We write the expression as 
\begin{equation}
S_{BLC}=\left(\frac{A_g}{A_{Pl}}\right)^{1+\frac{\delta}{2}}+\alpha \left(1+\frac{\delta}{2}\right)\ln \left(\frac{A_g}{A_{Pl}}\right)+\beta~~~~,    
\end{equation}
the subscript ``BLC" is implying the logarithmic corrected Barrow entropy. Here $\delta$, $\alpha$ and $\beta$ are model  parameters\cite{capozziello2025barrow}.
When the Schwarzschild BH's surface area is substituted, the entropy above takes the following form 
\begin{equation}\label{P4_S_BLC}
    S_{BLC} = (\pi r_{h}) ^{1+\frac{\delta }{2}} +\alpha \left\{\log (2 D+r_h)+\left(1+\frac{\delta
   }{2}\right) \log (\pi  r_h)\right\} + \beta ~~~~.
\end{equation}
Some other entropy corrections can be named in this stage. Non-extensive statistical mechanics generates the idea of Tsalli's entropy given by \cite{Rastgoo2020} $$S_T=\frac{A_g}{4}\left(\frac{A_g}{A_{Pl}}\right)^{1-q}~~~,~~q\in \mathbb{R}~~~~.$$ The core idea of this nonadditive modification of entropy is a generalisation of Shannon/Boltzmann entropy. This entropy is explicitly nonentensive for $q\neq 1$. 

Generalised information theory and idea of multifractal systems originates Renyi entropy given as \cite{Barrow1990} $$S_R=\frac{1}{\xi}\ln \left(1+\xi S_T\right)~~~~,$$ where $\xi\in \mathbb{R}$ is a constant. This model is used to reinterpret BH thermodynamics via information theory. This also generalizes Shannon/Boltzmann entropy.

Barrow entropy is considered better suited for thermodynamic studies in quantum gravity and  cosmology under certain contexts $-$ particularly when quantum gravitational or fractal space-time effects are expected to modify the classical laws of thermodynamics. From geometric origin, Barrow entropy is directly linked with horizon structure of the concerned BH. In thermodynamic study of gravity, entropy must reflect horizon geometry. The term $\Delta$ in the Barrow entropy represents quantum gravitational fluctuations, especially these causing dimensional deviations in the area of the horizon. Hence acting as a quantum corrected Bekenstein-Hawking law, Barrow entropy is a natural candidate to study semiclassical thermodynamics of BHs. Some forms of Renyi or Tsalli's entropy may violate the generalised second law of thermodynamics \cite{heydarzade2017black, ATLAS:2017gqz} under certain dynamics unless additional constraints are imposed. Barrow entropy preserves a monotonic, algebraic form of entropy \cite{ATLAS:2021kog} that scales with area in a modified way but does not introduce logarithmic (Renyi) or inverted nonlinear (Tsallis) complications. Barrow entropy makes temperature and heat flow derivations more tractable, especially in cosmological horizons. 

In the next section, different thermodynamic parameters for EMS BH given by (\ref{P4_lapse}) will be studied with the logarithmic corrected fractalised Barrow entropy given by (\ref{P4_S_BLC}).
%%%%%%%%%%%%%%%%%%%%%%%%%%%%%%%%%%%%%%%%%%%%%%%%%%%%%%%%%%
\section{Thermodynamic Quantities}\label{thermo}
%%%%%%%%%%%%%%%%%%%%%%%%%%%%%%%%%%%%%%%%%%%%%%%%%%%%%%%
From equation (\ref{P4_lapse}), for a vanishing lapse function, we obtain the mass as a function of $r_h$, the radius of the event horizon, as
\begin{equation}\label{P4_M}
    M = \frac{1}{12} \left(r_h \sqrt{24 \lambda  q^2+\lambda ^2
   r_h^4-6 \lambda  r_h^2+9}-\lambda  r_h^3+3 r_h\right)~~.
\end{equation}
\begin{table}[h]
\centering
\textbf{Table 1} \\[0.9em]
\begin{tabular}{|c|c|c|c|c|}
\hline
\multicolumn{2}{|c|}{\textbf{Parameters}} & \textbf{Hawking Temperature} & \textbf{Specific Heat} & \textbf{Gibbs Free energy} \\
\multicolumn{2}{|c|}{}&(Fig 1a - 1e)&(Fig 2a - 2e)&(Fig 3a - 3e \& 4)\\
\hline
 Black & q & $q_1 < q_2 \implies T_H(q_1)> T_H(q_2)$ & $q_1 < q_2 \implies$& $q_1 < q_2 \implies F(q_1) < F(q_2)$ \\
 Hole & & & $C_P(q_1)< C_P(q_2)$ & \\
\cline{2-5}
 Parameters & $\lambda$ & $\lambda_1 < \lambda_2 \implies T_H(\lambda_1) < T_H(\lambda_2)$ & $\lambda_1 < \lambda_2 \implies$ & For $S_4<S_{BLC}<S_5,$ \\
& & &$C_p(\lambda_1) > C_P(\lambda_2)$ & $\lambda_1 < \lambda_2 \implies F(\lambda_1) < F(\lambda_2)$ \\
\cline{5-5}
&&&&For $S_{BLC}>S_5,  \lambda_1 < \lambda_2 $ \\
&&&&$\implies F(\lambda_1) > F(\lambda_2)$ \\
\hline
  Barrow& & For $S_{BLC} < S_1,$ &  & For $S_{BLC} < S_3,~\delta_1 <\delta_2$ \\
  fractal& & $~\delta_1 <\delta_2 \implies T_H(\delta_1)>T_H(\delta_2)$ & & \\
  form&&&&$\implies F(\delta_1)<F(\delta_2)$\\
\cline{3-3}
  Parameter& $\delta$ &For $S_1<S_{BLC} < S_2,~$&$~\delta_1 <\delta_2 \implies$ & \\
 & & $\delta_1 <\delta_2 \implies T_H(\delta_1)<T_H(\delta_2)$ & & \\
&&& $C_P(\delta_1)>C_P(\delta_2)$ &\\
\cline{5-5}
&&&&For $S_{BLC} > S_3,~\delta_1 <\delta_2$  \\
&&&&$\implies F(\delta_1)>F(\delta_2)$\\
\cline{3-3}
&&For $S_{BLC} > S_2,~$&& \\
&&$\delta_1 <\delta_2 \implies T_H(\delta_1)>T_H(\delta_2)$ & & \\
\cline{1-5}
Logarithmic& $\alpha$ & $\alpha_1 < \alpha_2 \implies T_H(\alpha_1)>T_H(\alpha_2)$ & $\alpha_1 < \alpha_2 \implies$ & $\alpha_1 < \alpha_2 \implies F(\alpha_1)>F(\alpha_2)$  \\
corrected&&& $C_P(\alpha_1)>C_P(\alpha_2)$ & \\
\cline{2-5}
entropy& $\beta$ & $\beta_1 < \beta_2 \implies T_H(\beta_1) < T_H(\beta_2)$ & $\beta_1 < \beta_2 \implies$ & $\beta_1 < \beta_2 \implies F(\beta_1) > F(\beta_2)$ \\
Parameters&&& $C_P(\beta_1) > C_P(\beta_2)$ &\\
\hline
\end{tabular}
\\[0.9em] \textbf{Table 1: Parameter-wise effects on different thermodynamic quantities}
\end{table}

 In table 1, we have incorporated the effects of different model parameters contributed by the gravity theory and modification to entropy to the thermodynamic variables

When we dive into BH thermodynamics, we begin by looking at the BH's temperature. The BH's temperature has a clear definition in a general setting because its concept is geometric. This leads to the BH's temperature being expressed with the help of equation (\ref{P4_lapse}) as 
\begin{equation}\label{P4_T}
    T_H = \left.\frac{u'(r)}{4 \pi} \right|_{r_h}= \frac{1}{4 \pi} \left\{\frac{2
   M}{r_h^2}-\frac{1}{3} \lambda 
   \left(2r_h-\frac{q^2}{M}\right) \right\}~~~~.
\end{equation}
%%%%%%%%%%%%%%%%%%%%%%%%%%%%%%%%%%%%%%%%%%%%%%%%%%%%%%%%%%%%%%%%%%%%%%%%%%%%%%%%%%%%%%%%%%%%%
\begin{figure}[h!]
    \centering
     ~~~~~~~Fig ~1a ~~~~~~~~~~~~~~~~~~~~~~~~~~~~~~~~~~~~~~~~~~~~~~~~~~~~Fig~1b~~~~~~~\\
    \includegraphics[height = 2.2in, width=2.2in]{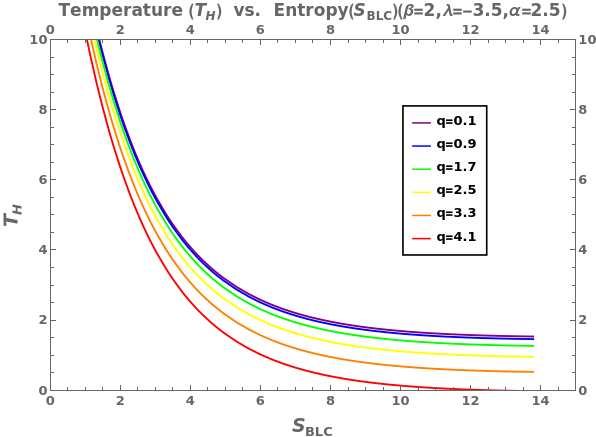}
    \includegraphics[height = 2.2in, width=2.2in]{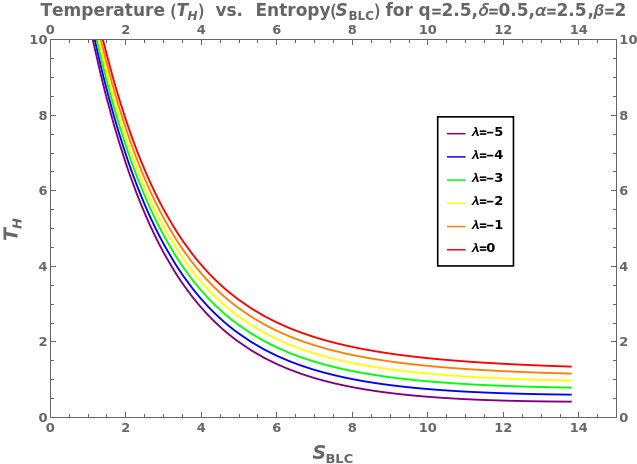}\\
    Figure : Fig 1a and 1b  are Hawking temperature(in $GeV$ unit) vs. logarithmic corrected Barrow entropy(dimensionless) plots for different values of $q$ and $\lambda$ keeping $M$ and $D$ fixed at $1$.
    \label{fig:temperature_heat}
\end{figure}
%%%%%%%%%%%%%%%%%%%%%%%%%%%%%%%%%%%%%%%%%%%%%%%%%%%%%%%%%%%%%%%%%%%%%%%%%%%%%%%%%%%%%%%%%%%%%
%%%%%%%%%%%%%%%%%%%%%%%%%%%%%%%%%%%%%%%%%%%%%%%%%%%%%%%%%%%%%%%%%%%%%%%%%%%%%%%%%%%%%%%%%%%%%
\begin{figure}[h!]
    \centering
      ~~~~~~~Fig ~1c ~~~~~~~~~~~~~~~~~~~~~~~~~~~Fig ~1d~~~~~~~~~~~~~~~~~~~~~~~~~Fig~1e~~~~~~~\\
      \includegraphics[height = 2.2in, width=2.2in]{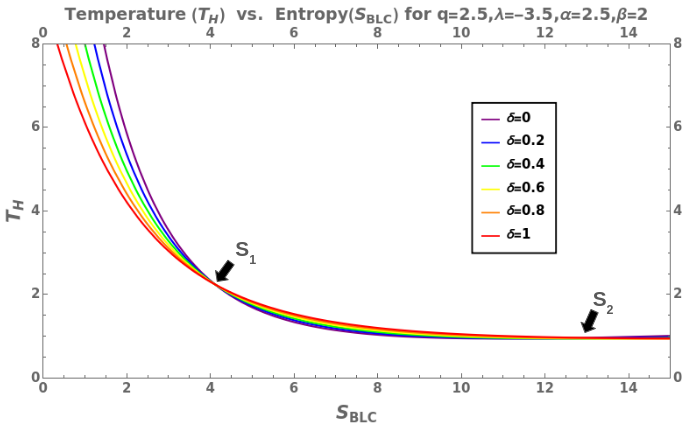}
      \includegraphics[height = 2.2in, width=2.2in]{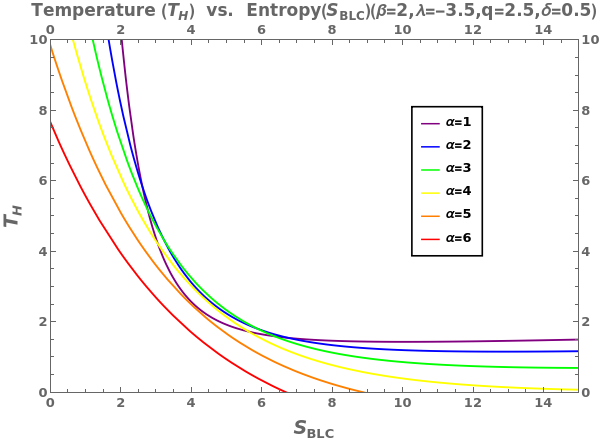}
        \includegraphics[height = 2.2in, width=2.2in]{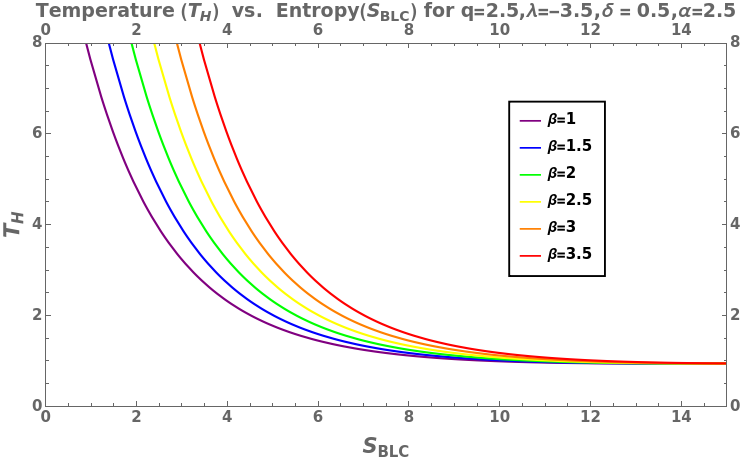}\\
    Figure : Fig 1c, 1d and 1e  are Hawking temperature(in $GeV$ unit) vs. logarithmic corrected Barrow entropy(dimensionless) plots for different values of $\delta, ~ \alpha$ and $\beta$ keeping $M$ and $D$ fixed at $1$.
\end{figure}
%%%%%%%%%%%%%%%%%%%%%%%%%%%%%%%%%%%%%%%%%%%%%%%%%%%%%%%%%%%%%%%%%%%%%%%%%%%%%%%%%%%%%%%%%%%%%
If we check for the units of different parameters and  $M(kg)$, unit of $T_H$ turns Kelvin. But here, in this work, we have chosen Planckian units ($\hbar=c=k\kappa_B=G=1$) and hence temperature is having the units of energy, e.g, $GeV$ or $eV$. 

In figure 1a to 1e, Hawking temperature $T_H$ is plotted against the logarithmically corrected Barrow's fractalised entropy. In 1a and 1b, $q$ and $\lambda$ are varied, 1c incorporates the variation of Barrow's fractalization index $\delta$ in the interval $[0,~1]$. Logarithmic correction parameters $\alpha$ and $\beta$ are varied in Fig. 1d and 1e, respectively.

In general, in all the cases, temperature is followed to decrease with the increment of entropy but temperature approaches a finite nonzero constant at large $S_{BLC}$, i.e., evaporation stops or freezes out at a minimal temperature $T_{min}$. 

In the works concerning quantum gravity effects, eg, inclusion of generalized uncertainty principle (GUP) \cite{adler2001generalized}, non extensive entropy (Tsallis, Barrow)\cite{li2022quasinormal} brings this kind of temperature variations. Popularly, this situation is named as having a ``minimal temperature plateau". 

An important information conservation happens while temperature plateaus at a minimum finite value. The remnant can store information, otherwise, a full evaporation will lead us to information loss. This offers a possible resolution to the BH information loss paradox. 

Some speculations are there where remnants can contribute to dark matter/ dark energy. BH remnants may behave as quantum field excitations with extremely high density and low temperature. Vacuum expectation value, i.e., the zero point energy of such fields could be nonzero and mimic a cosmological constant
$\rho_{vac}^{remnant}\sim \frac{1}{M^4}\sum_{i}<0|H_i|0>$, where the symbol  $H_i$ typically denotes the Hamiltonian operator for the $i$-th quantum field or mode in the theory.

This energy is effectively constant and does not dilute with the expansion of the universe. This is dark energy's behavior. The high entropy of remnants could contribute to the microstate of space time aligning with holographic dark energy models .

If remnants are sufficiently dilute, they do not conflict with cosmic microwave background radiation (CMB), big bang nucleosynthesys (BBN) or structure formation. Throughout gravitational vacuum energy can add up over cosmic scales \cite{adler2001generalized}. Temperature for very low $q$ and high $\alpha$ cases, however, vanishes inspite of ending at a finite plateau.

%%%%%%%%%%%%%%%%%%%%%%%%%%%%%%%%%%%%%%%%%%%%%%%%%%%%%%%%%%%%%%%%%%%%%%%%%%%%%%%%%%%%%%%%%%%%%

Thermodynamic principles tell us that heat capacity \cite{davies1978thermodynamics} shows how well a system can soak up thermal energy when its temperature goes up by one degree. We can figure out the specific heat $C_p$ by using equations (\ref{P4_T}) and (\ref{P4_S_BLC}) in this connection 
\begin{equation}\label{P4_C_P}
    C_P = \left(T_H \frac{\partial S_{BLC}}{\partial T
    _H} \right)_P = \frac{\left\{\alpha  \left(\frac{1}{2 D+r_h} + \frac{\frac{\delta}{2} + 1}{r_h}\right)+\pi ^{\frac{\delta}{2}+1} \left(\frac{\delta}{2}+1\right) r_h^{\delta /2}\right\}\left\{\frac{1}{3} \lambda  \left(r_h-\frac{q^2}{M}\right)-\frac{2 M}{r_h^2} + \frac{\lambda r_h}{3} \right\}} { 2 \left( \frac{\lambda r_h^3}{3} + \frac{M}{r_h^3} \right)}
\end{equation}
SI unit of specific heat is $JK^{-1}$. In natural units, it turns $GeV$. In figure 2a-2e, specific heat $C_P$ vs $S_{BLC}$ are plotted for varying $q$, $\lambda$, $\delta$, $\alpha$ and $\beta$ respectively. In fig 2a, $q=0.1,~0.9 ~ \text{and} ~ 1.7 $ and in fig 2b, $\lambda=0$ are different from the common trend followed by all other lines throughout these five figures. 

Let us discuss regarding the so called ``common feature" at first. For small and positive $S_{BLC}$, $C_P$ is firstly negative which turns suddenly positive after reaching a point $S_{BLC, 1}$. Again through an infinite jump, it turns negative (say at $S_{BLC, 2}$). Next, it changes its sign smoothly at $S_{BLC, 3}$. Overall four phases are noted. Unstable $\xrightarrow{S_{BLC,1}}$ Stable $\xrightarrow{S_{BLC,2}}$ Unstable $\xrightarrow{S_{BLC,3}}$ Stable. This nature is highly nontrivial. Before getting a dip into the concerned physics, let us review what happens when the ``trivial case", i.e., the classical BH thermodynamics is considered. 
The initial stage $0<S_{BLC}<S_{BLC, 1}$ is like the classical regime just discussed. At $S_{BLC, 1}$, the continuous change in the sign of the specific heat, without any divergence, suggests the BH undergoes a smooth thermodynamic crossover between stable and unstable regimes. This implies that the quantum corrections to BH thermodynamics modify evaporation gradually without introducing a singular phase boundary. The microstructure of concerned space-time is softly deforming here, possibly, interpolating between classical and quantum phases, i.e., a quantum-classical blended region. Thermodynamic instabilities are regularized smoothly.
We call the incident at $S_{BLC, 2}$ a phase transition. Below the critical point ($S_{BLC}<S_{BLC, 2}$), the BH behaves like a quantum stabilized object (Planck-scale remnant or regular curve). At $S_{BLC, 2}$, i.e., at the divergence, quantum gravity ``activates" and horizon fluctuates significantly. For $S_{BLC}>S_{BLC, 2}$, the system becomes semiclassical, classical evaporation dominates and instability returns. The infinite divergence in $C_P$ resembles a second order phase transition in standard thermodynamics.

From the equation (\ref{entropy}) and (\ref{temp}) $T\propto \frac{1}{M}$ and $c=\frac{dM}{dT}\propto -M^2<0$, i.e., the BH gets hotter as it looses mass, leading to runaway evaporation. 

%%%%%%%%%%%%%%%%%%%%%%%%%%%%%%%%%%%%%%%%%%%%%%%%%%%%%%%%%%%%%%%%%%%%%%%%%%%%%%%%%%%%%%%%%%%%%
\begin{figure}[h!]
    \centering
     ~~~~~~~Fig ~2a ~~~~~~~~~~~~~~~~~~~~~~~~~~~~~~~~~~~~~~~~~~~~~~~~~~~~Fig~2b~~~~~~~\\
    \includegraphics[height = 2.2in, width=2.2in]{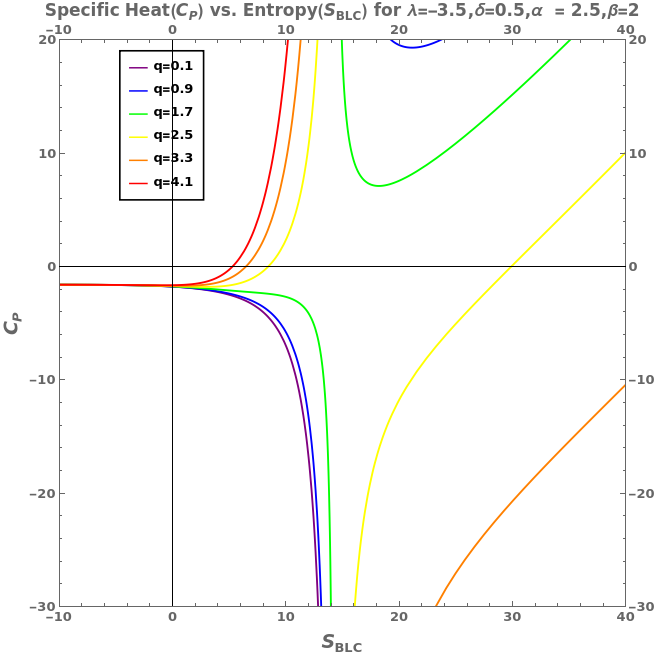}
    \includegraphics[height = 2.2in, width=2.2in]{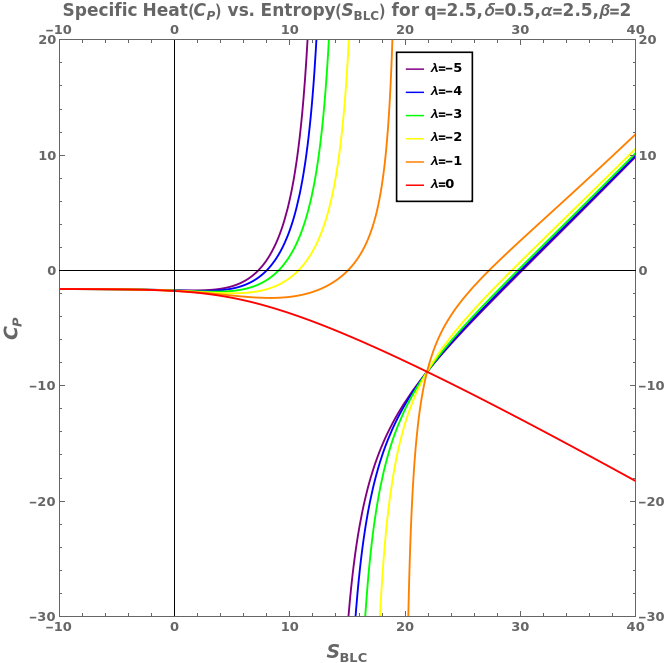}\\
    Figure : Fig 2a and 2b  are Specific heat(in GeV) vs. logarithmic corrected Barrow entropy(dimensionless) plots for different values of $q$ and $\lambda$ keeping $M$ and $D$ fixed at $1$.
    \label{fig:temperature_heat}
\end{figure}
%%%%%%%%%%%%%%%%%%%%%%%%%%%%%%%%%%%%%%%%%%%%%%%%%%%%%%%%%%%%%%%%%%%%%%%%%%%%%%%%%%%%%%%%%%%%%
%%%%%%%%%%%%%%%%%%%%%%%%%%%%%%%%%%%%%%%%%%%%%%%%%%%%%%%%%%%%%%%%%%%%%%%%%%%%%%%%%%%%%%%%%%%%%
\begin{figure}[h!]
    \centering
      ~~~~~~~Fig ~2c ~~~~~~~~~~~~~~~~~~~~~~~~~~~Fig ~2d~~~~~~~~~~~~~~~~~~~~~~~~~Fig~2e~~~~~~~\\
      \includegraphics[height = 2.2in, width=2.2in]{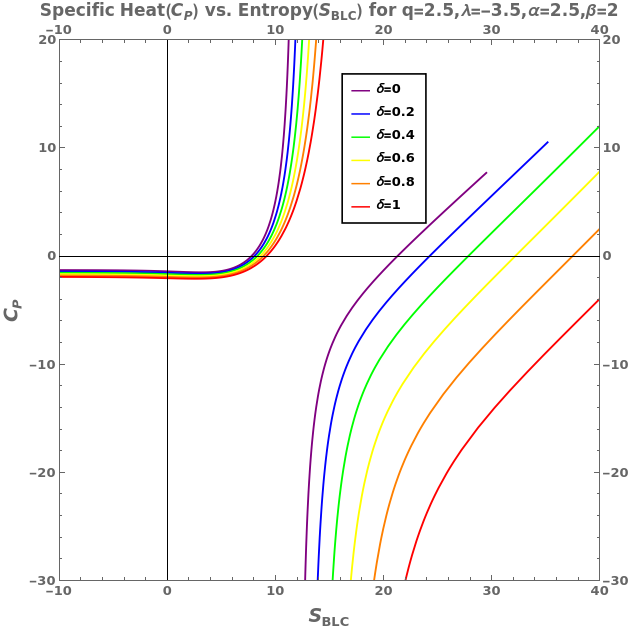}
      \includegraphics[height = 2.2in, width=2.2in]{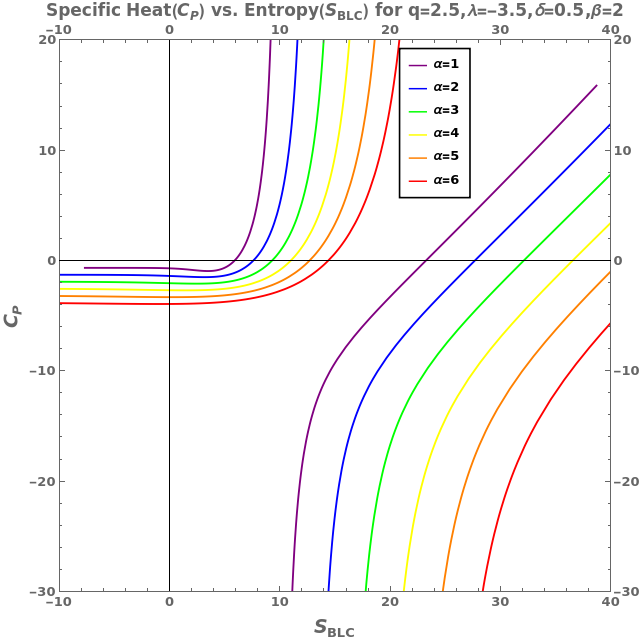}
        \includegraphics[height = 2.2in, width=2.2in]{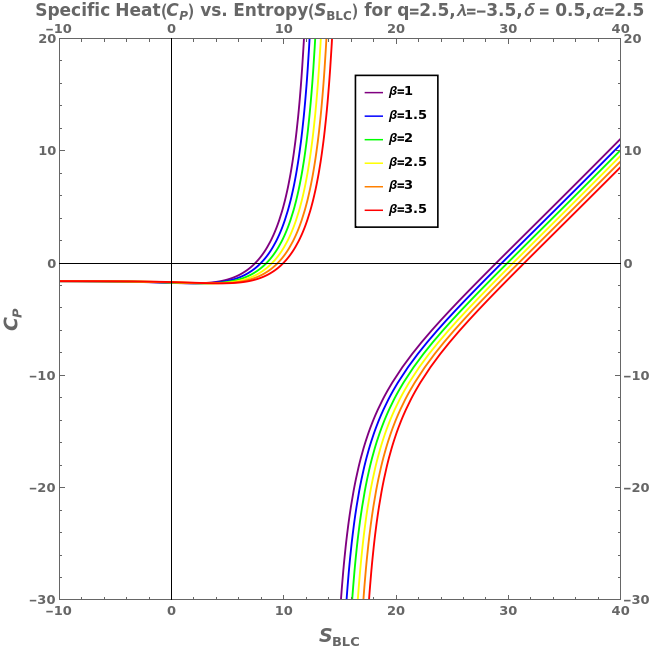}\\
    Figure : Fig 2c, 2d and 2e  are Specific heat(in GeV) vs. logarithmic corrected Barrow entropy(dimensionless) plots for different values of $\delta, ~ \alpha$ and $\beta$ keeping $M$ and $D$ fixed at $1$.
\end{figure}
%%%%%%%%%%%%%%%%%%%%%%%%%%%%%%%%%%%%%%%%%%%%%%%%%%%%%%%%%%%%%%%%%%%%%%%%%%%%%%%%%%%%%%%%%%%%%

The Gibbs free energy $(F)$ plays a key role in thermodynamics. It helps us to study phase changes in  BH systems \cite{kubizvnak2012p, gibbs1878equilibrium, tuck2019gibbs, middelburgthermodynamics}. We can write it as $F = M - T_HS_B$. It shows which phase might be more likely. So, we can express $F$ using the basic thermodynamic factors as
\begin{dmath}\label{P4_F}
    F = M - T_HS_{BLC} =\frac{1}{12} \left(r_h \sqrt{24 \lambda  q^2+\lambda ^2 r_h^4-6 \lambda  r_h^2+9}-\lambda  r_h^3+3 r_h\right)-\frac{1}{4 \pi }\left\{-\frac{1}{3} \lambda \left(r_h-\frac{q^2}{M}\right)+\frac{2 M}{r_h^2}-\frac{\lambda  r_h}{3}\right\} \left[\beta +\alpha  \left\{ \log (2 D+r_h)+\left(\frac{\delta}{2}+1\right) \log (\pi  r_h)\right\}+\pi ^{\frac{\delta }{2}+1} r_h^{\frac{\delta }{2}+1}\right]~~~~.
\end{dmath}
Unit of free energy is simply $J$. In natural units, it is simply $GeV$ or $eV$.
%%%%%%%%%%%%%%%%%%%%%%%%%%%%%%%%%%%%%%%%%%%%%%%%%%%%%%%%%%%%%%%%%%%%%%%%%%%%%%%%%%%%%%%%%%%%%
\begin{figure}[h!]
    \centering
     ~~~~~~~Fig ~3a ~~~~~~~~~~~~~~~~~~~~~~~~~~~~~~~~~~~~~~~~~~~~~~~~~~~~Fig~3b~~~~~~~\\
    \includegraphics[height = 2.2in, width=2.2in]{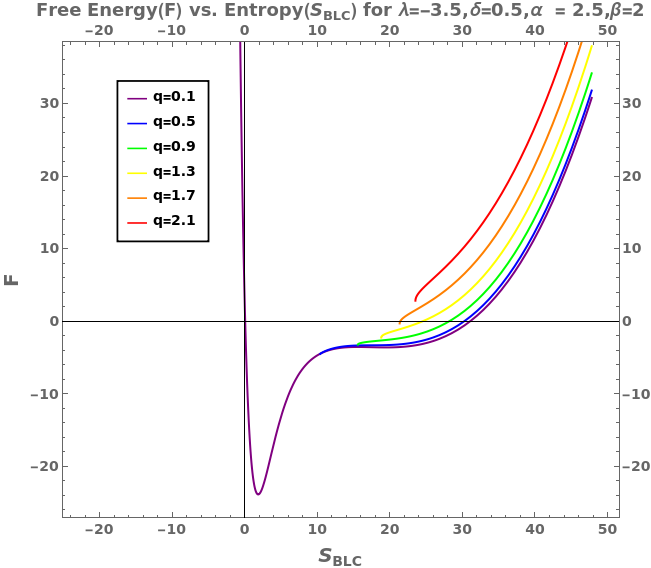}
    \includegraphics[height = 2.2in, width=2.2in]{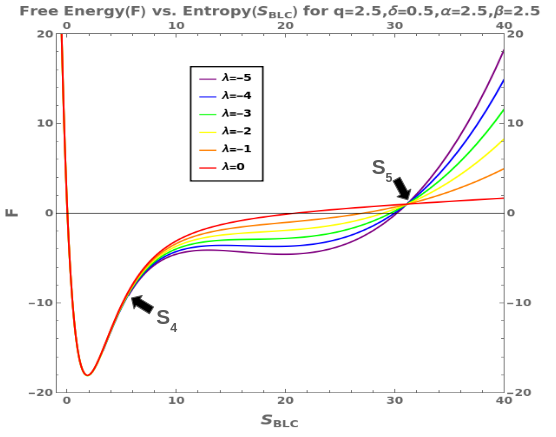}\\
    Figure : Fig 3a and 3b  are Free energy(GeV) vs. logarithmic corrected Barrow entropy(dimensionless) plots for different values of $q$ and $\lambda$ keeping $M$ and $D$ fixed at $1$.
    \label{fig:temperature_heat}
\end{figure}
%%%%%%%%%%%%%%%%%%%%%%%%%%%%%%%%%%%%%%%%%%%%%%%%%%%%%%%%%%%%%%%%%%%%%%%%%%%%%%%%%%%%%%%%%%%%%
%%%%%%%%%%%%%%%%%%%%%%%%%%%%%%%%%%%%%%%%%%%%%%%%%%%%%%%%%%%%%%%%%%%%%%%%%%%%%%%%%%%%%%%%%%%%%
\begin{figure}[h!]
    \centering
      ~~~~~~~Fig ~3c ~~~~~~~~~~~~~~~~~~~~~~~~~~~Fig ~3d~~~~~~~~~~~~~~~~~~~~~~~~~Fig~3e~~~~~~~\\
      \includegraphics[height = 2.2in, width=2.2in]{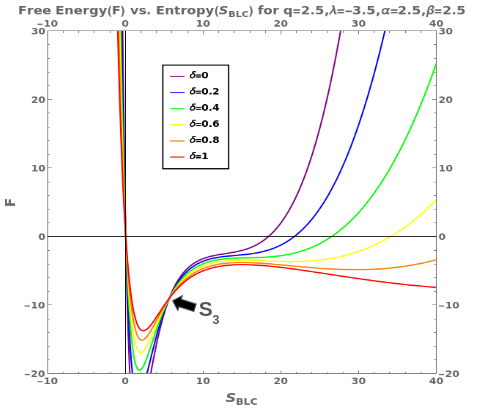}
      \includegraphics[height = 2.2in, width=2.2in]{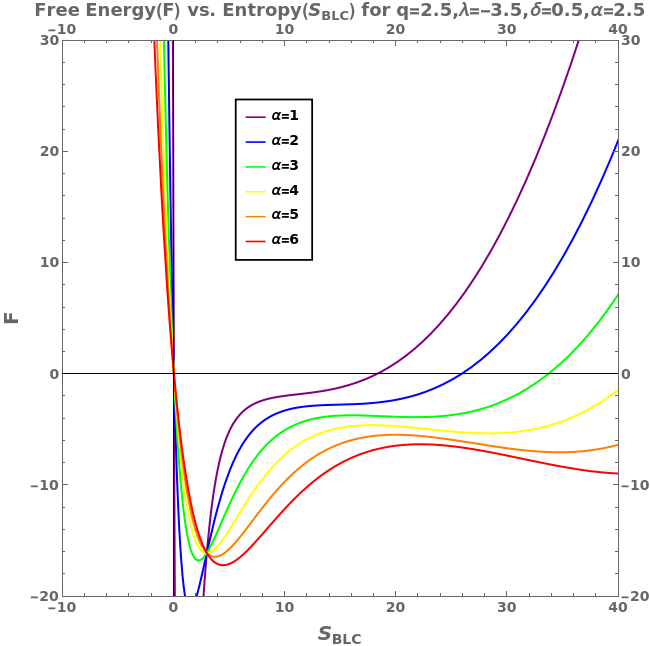}
        \includegraphics[height = 2.2in, width=2.2in]{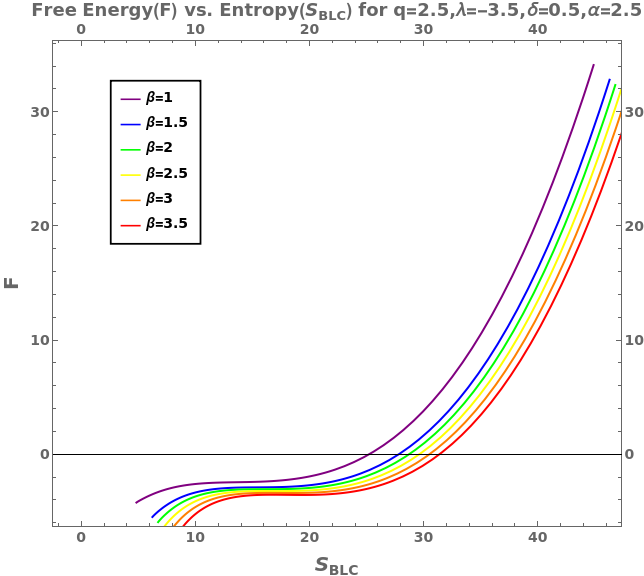}\\
    Figure : Fig 3c, 3d and 3e  are Free energy(GeV) vs. logarithmic corrected Barrow entropy(dimensionless) plots for different values of $\delta, ~ \alpha$ and $\beta$ keeping $M$ and $D$ fixed at $1$.
\end{figure}
%%%%%%%%%%%%%%%%%%%%%%%%%%%%%%%%%%%%%%%%%%%%%%%%%%%%%%%%%%%%%%%%%%%%%%%%%%%%%%%%%%%%%%%%%%%%%
%%%%%%%%%%%%%%%%%%%%%%%%%%%%%%%%%%%%%%%%%%%%%%%%%%%%%%%%%%%%%%%%%%%%%%%%%%%%%%%%%%%%%%%%%%%%%
\begin{figure}[h!]
    \centering
     ~~~~~~~Fig ~4~~~~~~~\\
    \includegraphics[height = 3in, width=3in]{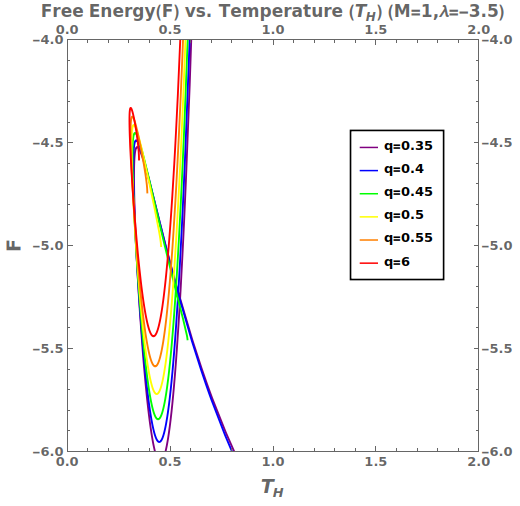}\\
    Figure Caption : Fig 4 is Free energy(GeV) vs. Hawking temperature(GeV) plot for $~M~ = ~1~, \lambda = -3.5, \delta = 0.5, D=1,~q=2.5$ and $\beta = 2$ .
    \label{fig:p_v_criticality}
\end{figure}
%%%%%%%%%%%%%%%%%%%%%%%%%%%%%%%%%%%%%%%%%%%%%%%%%%%%%%%%%%%%%%%%%%%%%%%%%%%%%%%%%%%%%%%%%%%%%
Free energy vs $S_{BLC}$ curves for varying $q$, $\lambda$, $\delta$, $\alpha$ and $\beta$ are plotted in figure 3a to 3e respectively. Except few exceptions at $\delta=1$ (fig 3c) and $\alpha=6$ (fig 3d), at small positive $S_{BLC}$, $F$ is positive and falls steeply. After turning negative, it possesses a minima and a local maxima and again a minima(or a point of inflection in place of the last two extrema) and then increases up to its positive values as we increase the value of $S_{BLC}$. If $F<0$, the BH is more thermodynamically favorable than the reference background. Free energy minimum in the negative zone implies $F(r_h)<0$ and $\frac{d^2F}{dr_h^2}>0$. This is thermodynamically performed over any configuration with higher $F$, such as thermal radiation or smaller/ larger BHs. The system will naturally relax into this configuration at a fixed temperature. As this is a global minimum, it dominates the canonical ensembles (i.e., most probable configuration at fixed temperature). Formation of a stable BH remnant is likely.
The part of the curves where either a maxima followed by a minima (or a point of inflection) are formed is signifying a metastable or unstable saddle point. Even though the configuration has a lower free energy than the reference background, it can not stay stable against small perturbations.

Free energy is plotted against $T_H$ in the figure 4. Only 
$q$ is varied. Overall trend is showing that free energy is double valued for maximum of the values that temperature can take. Nonmonotonicity of $T_H(r_h)$ and multivalued 
$F(T_H)$ allows the coexistence of phases. The crossing point of branches in $F(T_H)$ defines a first order phase transition temperature. The branch with the lowest free energy is thermodynamically favoured. The higher branch signifies negative specific heat and thus it is unstable (smaller BHs). Middle loop signifies metastable or unstable intermediate phase.

%%%%%%%%%%%%%%%%%%%%%%%%%%%%%%%%%%%%%%%%%%%%%%%%%%%%%%%%%%%%%%%%%%%%%%%%%%%%%%%%%%%%%%%%%%%%%
The BHs thermodynamics volume is given by 
\begin{equation}\label{P4_V}
V = \frac{4}{3} \pi  r^3~~~~.
\end{equation}
In natural units, the unit of of $V$ is similar to that of $Energy^{-3}$, i.e., $GeV^{-3}$.
Now substituting $\lambda = 3P$ in equation (\ref{P4_M}), we obtain the expression of the pressure as 
\begin{equation}\label{P4_P}
    P = \frac{M \left(2 M-4 \pi  r^2 T\right)}{r^2 \left(2 M
   r-q^2\right)}~~~~.
\end{equation} 
SI unit of $P$ is pascal. However, in natural units, it turns the dimensions of cosmological constant, $\Lambda\sim L^{-2}$ and hence the unit turns $GeV^4$.

%%%%%%%%%%%%%%%%%%%%%%%%%%%%%%%%%%%%%%%%%%%%%%%%%%%%%%%%%%%%%%%%%%%%%%%%%%%%%%%%%%%%
\section{P-V Criticality}\label{pv}
%%%%%%%%%%%%%%%%%%%%%%%%%%%%%%%%%%%%%%%%%%%%%%%%%%%%%%%%%%%%%%%%%%%%%%%%%%%%%%%%%%%%
Using the following equations, the critical point is determined as the P-V diagram's inflection point:
\begin{equation}\label{p4_pv_1}
    \left.\frac{\partial P}{\partial V} \right|_{T=T_c} = 0~~~~\text{and}
\end{equation}
and
\begin{equation}\label{P4_pv_2}
     \left.\frac{\partial^2 P}{\partial V^2}\right|_ {T=T_c}= 0~~~~.
\end{equation}
The ``critical pressure" $(P_c)$, ``critical volume" $(V_c)$, and "critical temperature" $(T_c)$ can be found by solving the equations (\ref{p4_pv_1}), (\ref{P4_P}), (\ref{P4_V}) and (\ref{P4_T}) as,
\begin{equation}\label{P4_V_C}
V_C = \frac{4 \pi }{3 \left(\frac{3 M r-q^2}{M r^3}\right)^{3/2}}~~,~~~~ T_C = \frac{3 M r-q^2}{2 \pi  r^3}~~\text{and}~~~~P_C = \frac{2 M \left(2 M r+q^2\right)}{r^3 \left(2 M r-q^2\right)}~~~~.
\end{equation}
%%%%%%%%%%%%%%%%%%%%%%%%%%%%%%%%%%%%%%%%%%%%%%%%%%%%%%%%%%%%%%%%%%%%%%%%%%%%%%%%%%%%%%%%%%%%%
\begin{figure}[h!]
    \centering
     ~~~~~~~Fig ~5~~~~~~~\\
    \includegraphics[height = 3in, width=3in]{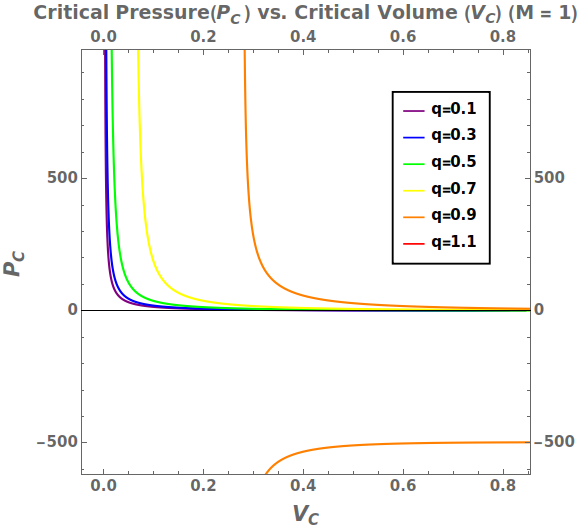}\\
    Figure Caption : Fig 5 is Critical Pressure($GeV^4$) vs. Critical Volume($GeV^{-3}$) plot for $~M~ = ~1~$ .
    \label{fig:p_v_criticality}
\end{figure}
%%%%%%%%%%%%%%%%%%%%%%%%%%%%%%%%%%%%%%%%%%%%%%%%%%%%%%%%%%%%%%%%%%%%%%%%%%%%%%%%%%%%%%%%%%%%%
We plot $P_c$ vs $V_c$ in the figure 5. We follow rectangular hyperbola type curves. This kind of relation reflects the scale invariance of the underlying gravitational theory at the critical point. This resembles the case of Vander Waal's gases. $P_c\sim \frac{1}{V_c^2}$ which is consistent with the BH horizon acting like a thermodynamic membrane. The space-time having a conformal symmetry near criticality. Since $V\propto r_h^3$ and $P\sim \frac{1}{r_h^2}$ which implies $P_cV_c^{\frac{2}{3}}=\text{constant}$. This suggests a dimensional balancing. This is nontrivial. In classical fluids, pressure and volume are extensive parameters. In this case, this emerge from sapce-time curvature. In a sense, the hyperbola charts the phase landscape of BH.

Again, by applying equation (\ref{P4_pv_2}) to the same, one can obtain
$$ V_C = \frac{8 \sqrt{2} \pi  M^{3/2}}{3 \left\{\frac{4 M^2 r^2+6 M r \left(2 M r-q^2\right)+5 \left(2 M r-q^2\right)^2}{r \left(4 M r-q^2\right)}\right\}^{3/2}}~,~~~
   T_C=\frac{4 M^2 r^2+6 M r \left(2 M r-q^2\right)+5 \left(2 M r-q^2\right)^2}{4 \pi  r \left(4 M r-q^2\right)}~,$$
\begin{equation}\label{P4_T_C_2}
   P_C = \frac{-4 M^3 r^3+6 M^2 r^2 \left(2 M r-q^2\right)+4 M^2 \left(4 M r-q^2\right)+5 M r \left(2 M r-q^2\right)^2}{r^2 \left(2 M r-q^2\right) \left(4 M r-q^2\right)}~~~~.
\end{equation}
There is no critical point without thermal fluctuations since it is evident from equation (\ref{P4_V_C}) and (\ref{P4_T_C_2}) that they do not satisfy one another simultaneously. Therefore, we must take into account the impact of thermal fluctuations in order to determine the critical points.

%%%%%%%%%%%%%%%%%%%%%%%%%%%%%%%%%%%%%%%%%%%%%%%%%%%%%%%%%%%%%%%%%%%%%%%%%%%%%%%%%%%%%%%%%%%%%
\section{Joule-Thomson Extension}\label{jt}
%%%%%%%%%%%%%%%%%%%%%%%%%%%%%%%%%%%%%%%%%%%%%%%%%%%%%%%%%%%%%%%%%%%%%%%%%%%%%%%%%%%%
The BH's electric potential can be obtained from the following thermodynamic equation,
\begin{equation}\label{P4_JT_U}
    U = \frac{2 \lambda  q r}{\sqrt{24 \lambda  q^2+\lambda ^2 r^4-6 \lambda  r^2+9}}~~~~.
\end{equation}
The JT expansion also known as the throttling process, takes place when gas moving from a high-pressure area to a low-pressure area through a porous plug. This process is not reversible. In this process, the enthalpy remains constant and the gas expands without transferring the heat. To determine the temperature change, we can use an isenthalpic process. The JT coefficient shows this temperature change in the throttling process. In other words, the JT coefficient is the numerical value of how steep an isenthalpic curve is on a $T-P$ diagram at any point. We use this symbol to represent it: 
\begin{equation}\label{P2_generalised_JT}
    \mu_{JT} = \left( \frac{\partial T}{\partial P} \right)_H ~~.
\end{equation}
 $\mu_{JT}$ is dimensionless while natural units are considered.

The set of all points where the JT coefficient equals zero is called the inversion curve. The JT coefficient has a key feature it divides the $T-P$ diagram into two areas. The area where $\mu > 0$ is known as the cooling region, while the area where $\mu < 0 $ is called the heating region. 

The JT coefficient for this BHs must be determined. This is obtained by first order differentiation of  the Smarr relation and then substituting it in the first law of BH thermodynamics, using $dH = dM = dq = 0$ and is given by, 
\begin{equation}\label{P2_JT_smarr}
    \mu_{JT} = \frac{1}{3S} \left[ 5V -2q\left( \frac{\partial U}{\partial P} \right)_H +2P\left( \frac{\partial V}{\partial P} \right)_H\right]~~.
\end{equation}
The equation of state for the  BH is given by the eq (\ref{P4_P}). Then inversion temperature takes the form
\begin{equation}\label{P4_T_i_d}
    T_i = V\left(\frac{\partial T}{\partial V} \right) = -\frac{r}{6 \pi } \left(P_i+\frac{2 M}{r^3}\right)~~~.
\end{equation}
Now with the help of the equations (\ref{P4_T}) and (\ref{P4_T_i_d}), substituting $\lambda = 3 P$, we can obtain the expression of inversion pressure as
\begin{equation}\label{P4_JT_P_i}
    P_i = \frac{10 M^2}{r^2 \left(4 M r-3 q^2\right)}~~~~.
\end{equation}
Substituting the eq. (\ref{P4_JT_P_i}) into (\ref{P4_T_i_d}), the equation of inversion temperature can be obtained as 
\begin{equation}\label{P4_JT_T_i}
    T_i = -\frac{r}{6 \pi } \left\{\frac{10 M^2}{r^2 \left(4 M r-3 q^2\right)}+\frac{2 M}{r^3}\right\}~~~~.
\end{equation}
%%%%%%%%%%%%%%%%%%%%%%%%%%%%%%%%%%%%%%%%%%%%%%
%%%%%%%%%%%%%%%%%%%%%%%%%%%%%%%%%%%%%%%%%%%%%%%
\begin{figure}[h!]
    \centering
     ~~~~~~~Fig ~6~~~~~~~~~~~~~~~~~~~~~~~~~~~~~~~~~~~~~~~~~~~~~~~~~~~~Fig~7~~~~~\\
    \includegraphics[height = 3in, width=3in]{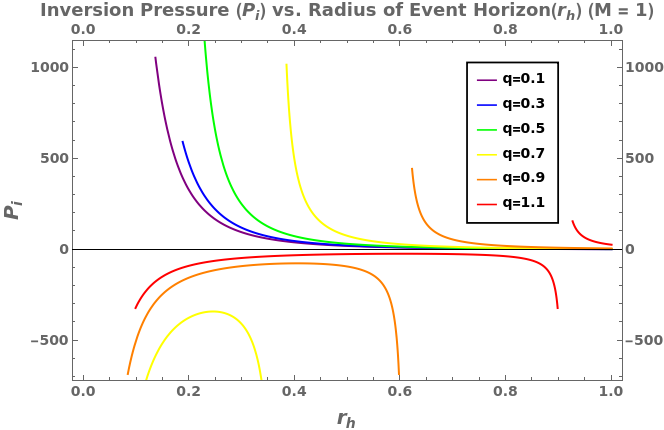}
    \includegraphics[height = 3in, width=3in]{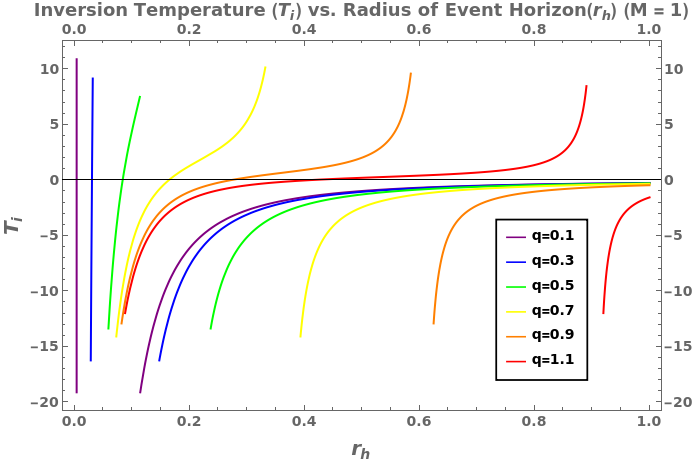}\\
    Figure Caption : Fig 6 and 7 are Inversion Pressure($GeV^4$) vs. radius of event horizon ($GeV^{-1}$) and Inversion Temperature($GeV$) vs. radius of event horizon plots respectively for $~M~ = ~1~$.
    \label{fig:J-T_INV_P_INV_T}
\end{figure} 
Inversion pressure and inversion temperature are plotted against the radius of event horizon in the figures 6 and 7 respectively for different values of $q$. Inversion pressure diverges infinitely and changes value from negative to positive. Inversion pressure diverges infinitely and changes value from negative to positive. Inversion temperature falls firstly as $r_h$ increases. Then it smoothly increases and smoothly changes value from negative to positive zone. For any BHs (eg., Reissner-Nordstrom AdS), temperature approaches zero at extremality $T_H\rightarrow 0$ as $r_h\rightarrow r_{ext}$. At this point, entropy is still finite, but the system cannot cool further. The JT expansion relics on the gradient $\left(\frac{\partial T_H}{\partial P}\right)_M$, which diverges if $T_H$ is flattening while $P$ keeps varying : the BH becomes incompressible from a thermal point of view. Further expansion can not cause meaningful cooling, because the BH is already at minimal possible temperature.

So the divergence of inversion pressure signals no inversion temperature exists below that point \cite{okcu2017joule}. In this article, a BH with richer phase structure, i.e., with Barrow entropy is considered. Disconnected branches of BH solutions are found with vertical discontinuity in $P_{inv}$, like $lim_{r_h \rightarrow r_c^-} P_{inv} \neq  lim_{r_h \rightarrow r_c^+} P_{inv}$. The system under goes a first order phase transition or a topological change to interrupt the continuity of the inversion process.

For certain parameter values (eg, minimal charge or horizon size), the BH temperature reaches a minimum or zero. When the curve reaches an extremum, $T_H\rightarrow 0 $ and $\mu_{JT}$ diverges. This causes a vertical asymptote or infinite jump in the inversion pressure as a function of entropy or volume. This also signals a phase transition between different BH geometries. It may reflect new degrees of freedom being activated or a topological shift. To bring analogy, we can look towards specific models like Gauss Bonnet or Lovelock gravity \cite{duval2022interacting}, the form of the inversion curve may lead to a pole in pressure.

The inversion temperature $T_{inv}$ in BH JT expansion is the temperature at which the cooling heating behavior reverses under an isenthalpic (constant mass) process. Above $T_{inv}$, expanding BH cools down (like real gas). Mathematically, a continuous, smooth curve suggests the existence of a $r_h=r_{h0}$ such that $T_{inv}(r_{h0})=0$ signifying a minimally sized remnant or a purely mathematical transition point. 

In standard thermodynamics, a negative inversion temperature region is unphysical. But a BH is in a regime where no cooling is possible, even under expansion. This is often interpreted as a quantum gravity regime or a signal of a thermodynamically dead remnant. As the BH grows, the space-time becomes more ``classical". Barrows fractal correction smooths out the inversion behavior. Barrow entropy models the horizon as a nonsmooth fractal surface. This causes entropy to saturate or grow slowly as area shrinks which prevents the run away evaporation.

%%%%%%%%%%%%%%%%%%%%%%%%%%%%%%%%%%%%%%%%%%%%%%%%%%%%%%%%%%%%%%%%%%%%%%%%%%%%%%%%%%%%%%%%%%%%%
According to the first law of BH thermodynamics, we can express the change in mass as follows: 
\begin{equation}\label{P4_mass_change}
    dM = Tds + \Phi dq + VdP~~,
\end{equation}
where $\Phi$ indicates the electric potential of the BH at the horizon
\begin{equation}\label{P4_ep}
    \Phi = \left( \frac{\partial M}{\partial q} \right)_{S,P}~ = ~~ \frac{q}{r_h}~~.
\end{equation}
for the constants $S$ and $P$. Considering $dM =dq=0$~ in the throttling process, we can write
\begin{equation}\label{P4_Gen_tem}
    T \left( \frac{\partial S}{\partial P} \right)_M + V = 0~~.
\end{equation}
where using thermodynamic variation ~ $dS = \left( \frac{\partial S}{\partial S} \right)_T dP + \left( \frac{\partial S}{\partial T} \right)_P dT$ ~ and Maxwell relation ~$\left( \frac{\partial S}{\partial P} \right)_T = - \left( \frac{\partial V}{\partial T} \right)_P$ ~ we obtain
\begin{equation}\label{p4_exp_JT}
    \mu_{JT} = \frac{1}{C_P} \left[ T\left( \frac{\partial V}{\partial T} \right)_P - V \right]~~,
\end{equation}
Using eqs. (\ref{P4_T}) and (\ref{P4_V}), one can obtain JT coefficient $\mu_{JT}$ from the eq. (\ref{p4_exp_JT}) as,
\begin{equation}\label{P4_JT_MU}
    \mu_{JT} = \frac{36 \pi  r^2 \left\{\frac{1}{3} \lambda \left(r-\frac{q^2}{M}\right)-\frac{2 M}{r^2}+\frac{\lambda  r}{3}\right\}-8 \pi  \left(6 M+\lambda  r^3\right)}{9 \left\{\alpha \left(\frac{1}{2 D+r}+\frac{\frac{\delta}{2}+1}{r}\right)+\pi ^{\frac{\delta }{2}+1} \left(\frac{\delta }{2}+1\right) r^{\delta /2}\right\} \left\{\frac{1}{3} \lambda \left(r-\frac{q^2}{M}\right)-\frac{2 M}{r^2}+\frac{\lambda  r}{3}\right\}}~~~.
\end{equation}

%%%%%%%%%%%%%%%%%%%%%%%%%%%%%%%%%%%%%%%%%%%%%%%%%%%%%%%%%%%%%%%%%%%%%%%%%%%%%%%%%%%%%%%%%%%%%
\begin{figure}[h!]
    \centering
     ~~~~~~~Fig ~8~~~~~~~\\
    \includegraphics[height = 3in, width=3in]{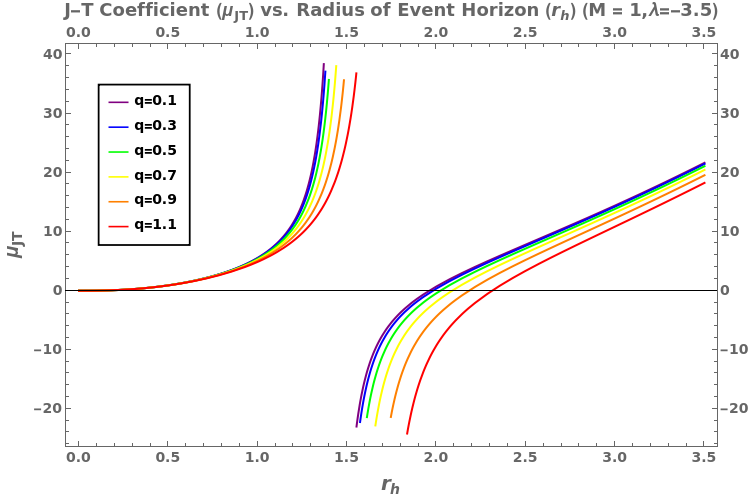}\\
    Figure Caption : Fig 8 is J-T Coefficient $(\mu_{JT})$(dimensionless) vs. radius of event horizon($GeV^{-1}$) plot $~M, ~\delta, ~D, ~\alpha~$ and $~\lambda$~ fixed at $1, ~0.5,~ 1,~10~$ and -3.5 respectively..
    \label{fig:J-T_MU}
\end{figure}
%%%%%%%%%%%%%%%%%%%%%%%%%%%%%%%%%%%%%%%%%%%%%%%%%%%%%%%%%%%%%%%%%%%%%%%%%%%%%%%%%%%%%%%%%%%%%
In figure 8, we plot JT coefficient with respect to the changes in the radius of event horizon. A vertical asymptote is followed to get constructed which force it to transit from a positively valued range to a negative one. One reason we suspect to be the fractal horizon isometry and phase decoupling. Barrow entropy implies the BH horizon becomes fractal at Planck scales. It has noninteger effective dimension. This affects thermodynamic volumes and pressure response, leading to disconnected thermodynamic phrases separated by $r_h=r_{h,0}$. The JT coefficient's divergence marks a decoupling of the thermodynamic behavior across this geometric transition. So we can conclude that the BH transits from a quantum dominated, fractional dimensional phase (small $r_h$) to a smooth classical manifold(large $r_h$).

The vertical asymptote is the critical value for the radius of event horizon where space time microstructure reorganizes.

Also in figure 8, JT coefficient shows a jump. A jump in the JT coefficient $\mu_{JT}$ for BHs signals a significant thermodynamic change, often a phase transition or a non-smooth inversion behavior. In the reference \cite{chabab2020excited}, authors found that the Joule–Thomson coefficient changes sign across inversion curves, with critical behavior near transition points. \cite{d2022forecasting}  observed a ``jump" or sudden change in the slope of inversion curves. \cite{okcu2017joule} computed $\mu_{JT}$ and inversion curves for various AdS BHs, observing discontinuities near phase transitions.

In the geometric thermodynamic framework, the thermodynamic curvature ${\cal R}$ should diverge at this vertical asymptote. This signifies strong correlations between BH microstates. The point where $\mu_{JT}\rightarrow \infty$ corresponds to a critical thermodynamic curvature singularity. This prediction motivates us to find singularities/ divergences in the Ricci curvature of the geometrothermodynamic space of this model, incorporated in the section \ref{GT}.
%%%%%%%%%%%%%%%%%%%%%%%%%%%%%%%%%%%%%%%%%%%%%%%%%%%%%%%%%%%%%%%%%%%%%%%%%%%%%%%%%%%%%%%%%%%%%
\section{Geometrothermodynamics}\label{GT}
%%%%%%%%%%%%%%%%%%%%%%%%%%%%%%%%%%%%%%%%%%%%%%%%%%%%%%%%%%%%%%%%%%%%%%%%%%%%%%%%%%%%
Here we evaluate the Weinhold metric\cite{weinhold1975metric, weinhold1975metric_II} for the thermodynamic space of this BH. The concerned Ruppeiner metric\cite{ruppeiner1995riemannian} can be computed by dividing it by the BH's temperature, $T_H$ at the event horizon. We obtain it into the form, given by the equations (\ref{R_1}), (\ref{R_2}) and (\ref{R_3}). After huge calculation, we derive the Ricci scalar for this metric and plot this with respect to $r_h$ and $q$ shown in the figure 8a - 8c for $~M = 1,~\lambda = -3.5,~ d = 1, ~\alpha = 10, ~\beta = 1$ and $\delta = 0.1,~                                                 0.5,$ and $0.9$ respectively.

\begin{equation}\label{R_1}
\frac{\partial^2 M}{\partial r_h^2}  =\frac{24 \pi  \lambda  M r_h^3 \left(\lambda  r_h^2-3\right)^2}{\left\{24 \lambda  q^2+\left(\lambda  r_h^2-3\right)^2\right\}^{3/2} \left(6 M^2-2 \lambda  M r_h^3+\lambda  q^2 r_h^2\right)}~~~.
\end{equation}

\begin{dmath}\label{R_2}
     \frac{\partial^2 M}{\partial r_h \partial q}  = \frac{12 \pi}{\frac{\lambda  q^2}{M}+\frac{6 M}{r^2}-2 \lambda  r}\left[-\frac{2 \lambda  r \left(\lambda  r^2-3\right) \left\{24 \lambda  q^2-3 \left(\lambda  r^2-1\right) \left(\sqrt{24 \lambda  q^2+\left(\lambda  r^2-3\right)^2}-\lambda 
   r^2+3\right)\right\}}{3 \left\{24 \lambda  q^2+\left(\lambda  r^2-3\right)^2\right\}^{3/2} \left\{\frac{2 \alpha }{2 d+r}+\pi ^{\frac{\delta }{2}+1} (\delta +2) r^{\delta /2}+\frac{\alpha  (\delta
   +2)}{r}\right\}^2}-\frac{\left\{24 \lambda  q^2-3 \left(\lambda  r^2-1\right) \left(\sqrt{24 \lambda  q^2+\left(\lambda  r^2-3\right)^2}-\lambda  r^2+3\right)\right\} \left(-\frac{2 \alpha }{(2
   d+r)^2}+\frac{1}{2} \pi ^{\frac{\delta }{2}+1} \delta  (\delta +2) r^{\frac{\delta }{2}-1}-\frac{\alpha  (\delta +2)}{r^2}\right)}{3 \sqrt{24 \lambda  q^2+\left(\lambda  r^2-3\right)^2} \left(\frac{2
   \alpha }{2 d+r}+\pi ^{\frac{\delta }{2}+1} (\delta +2) r^{\delta /2}+\frac{\alpha  (\delta +2)}{r}\right)^3}-\frac{4 \lambda  r^3 (2 d+r)^2 \left\{12 \lambda  q^2-\left(\lambda  r^2-2\right)
   \left(\sqrt{24 \lambda  q^2+\left(\lambda  r^2-3\right)^2}-\lambda  r^2+3\right)\right\}}{\left\{24 \lambda  q^2+\left(\lambda  r^2-3\right)^2\right\} \left\{2 d (\delta +2) \left(\alpha +\pi ^{\frac{\delta
   }{2}+1} r^{\frac{\delta }{2}+1}\right)+\pi ^{\frac{\delta }{2}+1} (\delta +2) r^{\frac{\delta }{2}+2}+\alpha  (\delta +4) r\right\}^2}\right]~~~.
\end{dmath}
\begin{dmath}\label{R_3}
    \frac{\partial^2 M}{\partial q^2}  = \\
   \frac{48 \pi  \lambda  M q r^3 (2 d+r) \left(24 \lambda  q^2-\lambda ^2 r^4+9\right)}{\left(24 \lambda  q^2+\left(\lambda  r^2-3\right)^2\right)^{3/2} \left(2 d (\delta +2) \left(\alpha +\pi ^{\frac{\delta
   }{2}+1} r^{\frac{\delta }{2}+1}\right)+\pi ^{\frac{\delta }{2}+1} (\delta +2) r^{\frac{\delta }{2}+2}+\alpha  (\delta +4) r\right) \left(6 M^2-2 \lambda  M r^3+\lambda  q^2 r^2\right)}~~~.
\end{dmath}
%%%%%%%%%%%%%%%%%%%%%%%%%%%%%%%%%%%%%%%%%%%%%%%%%%%%%%%%%%%%%%%%%%%%%%%%%%%%%%%%%%%%%%%%%%%%%
\begin{figure}[h!]
    \centering
      ~~~~~~~Fig ~9a ~~~~~~~~~~~~~~~~~~~~~~~~~~~Fig ~9b~~~~~~~~~~~~~~~~~~~~~~~~~Fig~9c~~~~~~~\\
      \includegraphics[height = 2.2in, width=2.2in]{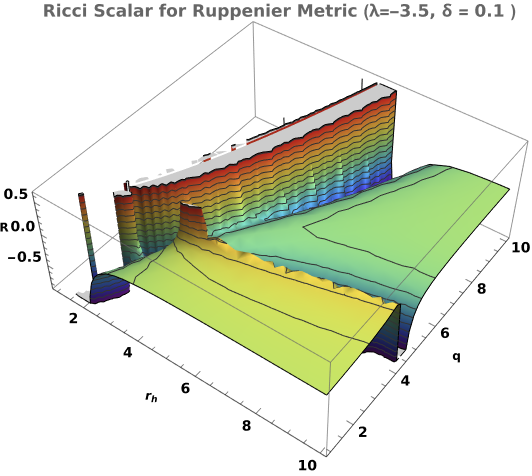}
      \includegraphics[height = 2.2in, width=2.2in]{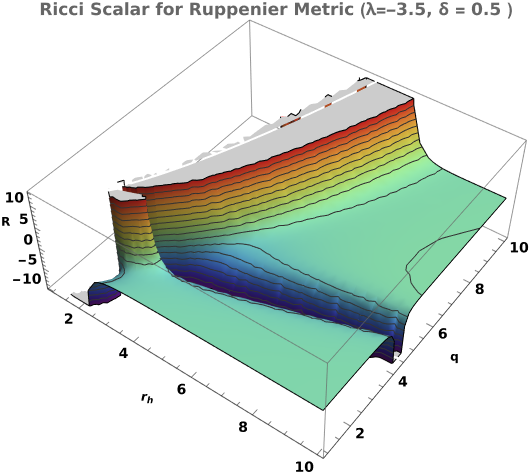}
        \includegraphics[height = 2.2in, width=2.2in]{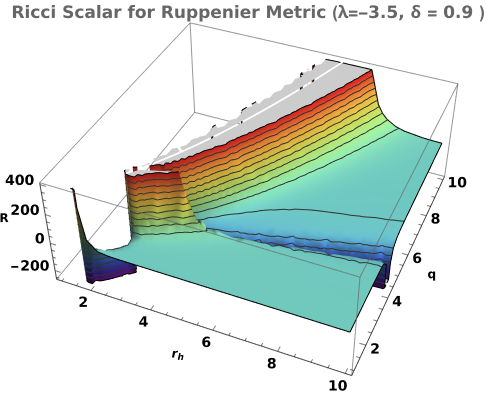}\\
    Figure : Fig 9a, 9b and 9c are the Ricci scalar plots w.r.t the radius of the event horizon $r_h$ ($GeV^{-1}$)  and BH charge $q$ (dimensionless) for $~M = 1,\lambda = -3.5, \d = 1, \alpha = 10, \beta = 1$ and $\delta = 0.1, 0.5,$ and $0.9$ respectively.
\end{figure}
%%%%%%%%%%%%%%%%%%%%%%%%%%%%%%%%%%%%%%%%%%%%%%%%%%%%%%%%%%%%%%%%%%%%%%%%%%%%%%%%%%%%%%%%%%%%%
We plot the Ricci curvature of the geometric space, ${\cal R}$ in the fig 9a-c for $\delta=0.1, ~0.5, ~0.9$ respectively. We follow two surfaces working as asymptotic ones along one $q$ constant and almost a $r$ constant  directions. As $\delta$ increased, the latter one immediately vanishes. For larger $\delta$, even the first one starts to vanish. So Barrow fractalised surface barely allows the geometrothermodynamic Ricci to vanish at high $r_h$ and high $q$.
%%%%%%%%%%%%%%%%%%%%%%%%%%%%%%%%%%%%%%%%%%%%%%%%%%%%%%%%%%%%%%%%%%%%%%%%%%%%%%%%%%%%%%%%%%%%%
\section{Brief Discussion and Conclusion}\label{conclusion}
%%%%%%%%%%%%%%%%%%%%%%%%%%%%%%%%%%%%
Barrow introduced the idea that a black hole's horizon may not be perfectly smooth. Rather at small (Planck) scales, it may become fractal-like. This changes the entropy area relation $S_{Beken}=\frac{A_g}{A_Pl}$ to $S_B=\left(\frac{A_g}{A_Pl}\right)^{1+\frac{\Delta}{2}}$. For $\Delta>0$, entropy grows faster than area. This points a nonlinear growth, i.e., small increase in mass yields larger entropy. Classical assumptions of local flatness at the horizon break down. Geometry is no longer smooth and hence heat-flow, radiation and curvature near the horizon get modified. Thermodynamic conjugate quantities shift, eg, $T_H=\frac{\partial S}{\partial M}$ changes. In one phrase mere degrees of freedom is enjoyed.

The logarithmic correction which arises from quantum statistical effects, lets the $log A_g$ term to dominate at small areas and hence entropy no longer vanishes as $A_g\rightarrow 0$. This slows the entropy loss in late-stage evaporation. Temperature is flattened and reduces to a finite minimum remnant like temperature. We expected, in this work, that the combo of fractalised entropy and logarithmic correction may give minimum temperature, stable remnants, modified phase structure and some new critical behaviors. This leads us to choose a black hole sitting in a gravity theory where gravity, electromagnetic field and a scalar field (real or complex) are all coupled. Classical no hair theorem is violated by the introduction of a spontaneous scalarization by activating nontrivially which again alters the entropy, temperature and phase structure. Scalarised Einstein-Maxwell-scalar black hole, are followed to emit distinct gravitational wave signatures. These are followed to be constructed and detected by LIGO/ VIRGO observations\cite{Doneva2018}.

Our motivation is to follow the thermodynamics of the quantized black hole of Einstein Maxwell scalar gravity with the quantum mechanical Barrow entropy model equipped with a logarithmic correction term. This study abides by the generalized second law under certain dynamics. 

In table 1, we have noted changes (and the conditions incorporated) for important thermodynamic variables with respect to different model parameters. In fig 1a-e, we have plotted Hawking temperature as a function of the logarithmic corrected entropy $S_{BLC}$. We follow the temperature to fall and maintain a finite minimum value once $S_{BLC}$ is large. This nature is contrary to the standard Hawking radiation where temperature diverges. Alike many other quantum gravity models (like GUP, LQG noncommutative geometry) black hole evaporation halts at a finite mass and size. The leftover object is a stable, Planck scale black hole remnant, with very high entropy, low temperature which possesses no further evaporation.

Recent observations of supermassive black holes show their mass increases at rate proportional to the cube of the cosmic scale factor $a^3$, independent from traditional mechanisms like accretion or mergers \cite{Farrah2023}, The observed coupling parameter $k\approx 3$ (with zero coupling ruled out at $99.98\%$ confidence) implies that $M_{BH}\propto a^k\approx a^3$. Meanwhile, number density of black holes falls as $a^{-3}$. Combined this, yields a constant total energy density in BHs $\rho_{BH}=\frac{n_{BH}M_{BH}}{a^3}$. Energy conservation in FLRW metric implies a negative pressure, i.e., behavior identical to a cosmological constant. Production of stellar remnant black holes across cosmic history matches the observed value of $\Omega_{\Lambda}\approx 0.68$\cite{Farrah2023}.

Next physical quantity is specific heat which is plotted in the fig 2a-e. Basic trend is to have small black holes with negative specific heat which changes the sign continuously. Then crossing a vertical asymptote another sign change is occurred. When the black holes grow larger, its specific heat turns positive again by a smooth crossing. Self gravitating systems (like star clusters or black holes) expands and cools when energy is added. This is the Jeans instability. Energy loss causes contraction and that increases temperature. At very small masses (Planck scale), we expect quantum gravitational effects to intervene. Barrow entropy flattens $T_H$, i.e., in spite of letting $T_H\rightarrow \infty$ we get $T\rightarrow  T_{min}$. Evaporation slows down at small mass and specific heat becomes zero or positive. This stabilizes the black hole to leave behind a black hole remnant. Quantum modification by Barrow indexing near a critical mass, quantum repulsion etc cancels gravitational pull and system turns thermodynamically stable with positive $C_P$. Inclusion of only Barrow entropy, on a gross, makes specific heat $\propto M^{1+\Delta}$. Simply, specific heat remarks negative and becomes more steeply negative as $\Delta$ increases. When logarithmic correction is involved, log term imposes quantum  cutoff affects at small scales.

The smooth-sign-change resembles a second order phase transition (like Hawking Page transition), where black holes ``condense" into a stable state. If $\beta<0$, the logarithmic correction completes destructively with the Barrow term. Hence at $S_{BLC,2}$, the curvature flips, causing $C_P$ to diverge. Finally, a large black hole is transitiing from negative specific heat to a positive specific heat zone. For $S_{BLC}>>S_{BLC,3}$, $C_P>0$ interprets the black hole can thermally colaborate with its environment (like AntideSitter or Reissner Nordstrom black holes). Unlike classical Schwarzschild case, quantum gravitational effects (Barrow+log) are found to stabilize large black holes. This smooth sign change resembles the Hawking-Page transition, where AntideSitter black holes become thermodynamically preferred at higher masses. Here, such a transition is driven by quantum corrections (not just by AntideSitter curvature).

We have plotted the characteristics of free energy in the figure 3a-e. Free energy attains a minima in the negative zone and then forms consequtive maxima and minima (or a point of inflection in stead) and then raises to positive values. Negative free energy is pointing towards a thermodynamically favored state or attractor. This also signifies stable condensation of quantum gravitational degrees of freedom on the horizon. These stable non evaporating remnants avoid the uncontrolled rise of temperature and instead from long lived or permanent objects. This supports models addressing the information paradox, dark matter coordinates and cosmic evolution. As the black hole grows larger, free energy changes sign from $F<0$ to $F>0$ which is a signature of a metastable remnant phase bounded by classical black hole behavior. 

In figure 4, free energy is multiple valued when is plotted against temperature. One branch represents a long-lived or stable remnant. The rest one is an unstable evaporating black hole. As $T_H$ decreases in the late evaporation stage, the black hole can settle onto the stable, lower $F$ remnant branch. Transition between these branches, as temperature or mass evolve, illastrates a rich, multi-phased thermodynamic structure, providing potential pathways for evaporation halting, remnant formation and even horizon topology tunneling in quantum gravity scenario \cite{Kim2016}.

During $P-V$ criticality analysis, we found, in figure 5, that the critical pressure vs critical volume show a Boyle's law like rectangular hyperbolic nature. Even with fractal modifications and loop corrections, critical behavior mirrors the classical Vander Waal's like pattern seen in Schwarzschild and Reissner Nordstrom AdS black holes. The hyperbolic relation shows that key phase transition structures survive under quantum gravitational corrections.

Figure 6 and 7 are respectively for inversion pressure and inversion temperature. Inversion pressure is followed to possess a vertical asymptote which highlights a unique feature indeed by quantum gravity corrections. Existence of a size threshold below which the black hole is in a strictly heating regime, consistent with the presence of a thermodynamically stable remnant. Inversion temperature, on the other hand, smoothly changes sign. This confirms the inversion temperature is a continuous function of horizon radius or pressure. It smoothly transits through zero, making the cooling heating boundary without discontinuities. This behavior is consistent with the weldefined Joule-Thomson expansion in this quantum gravity corrected framework. We have plotted Joule-Thomson coefficient vs radius of event horizon in figure 8. Whenever the Joule-Thomson coefficient exhibits a vertical asymptote, it is revealing a thermodynamic end point (the extremal, zero temperature regime). For the scenario where both the Barrow and log corrections are involved, this means if the black hole shrinks towards its quantum corrected extremal radius, $T_{inv}\rightarrow 0$ and $\frac{\partial T_{inv}}{\partial P_{inv}}\rightarrow \infty$ causing the divergence of $\mu_{JT}$. This signals an insurmountable boundary for cooling/heating process. Hence no further Joule-Thomson inversion can occur. This divergence elegantly marks the shift from dynamic evaporation into a thermodynamically frozen remnant. We find similar kind of outcomes in the articles \cite{javed2024joule, Das2024JouleThomson}.

Figure 9a-c are plotted to show the variation of the geometrothermodynamic Ricci curvature with respect to the radius of event horizon and the parameter $q$ for different values of the Barrow index $\delta$. According to geometrothermodynamics, a divergence in the Ricci scalar marks a breakdown of the equilibrium thermodynamic manifold. This signals a critical (second ordered) phase transition or a boundary in physical states. Barrow and logarithmic corrections modify entropy and temperature profoundly. When their interplay triggers instability, the Ricci scalar diverges. This shows one onset of rapid changes in heat capacity, partition function behavior or free energy slopes. This indicates the system is transiting between regimes, eg, from unstable Barrow dominant behavior to a stabilized quantum remnant.

In a nutshell, if we opt a quantum corrected gravity bypassing the no-hair theorem accompanied with fractally deformated entropy and logarithmic correction which involves correction upto quantum partitions, a thorough thermodynamic study leads us to the formation of high entropy, low temperature (plateau) stabilized quantum remnant. Formation of such primordial black holes in early universe, acting as a dark energy candidate, without requiring any extra field, may dissolve the ambiguities related to the evolution of dark matter dark energy.
\\

%%%%%%%%%%%%%%%%%%%%%%%%%%%%%%%%%%%%%%%%%%%%%%%%%%%%%%%%%%%%%%%%%%%%%
{\bf Acknowledgment : } RB thanks IUCAA, Pune for granting Visiting Associateship. SP thanks Department of Mathematics, The University of Burdwan for different research facilities.

\section*{Data Availability Statement}
My manuscript has no associated data. [Authors’ comment: The authors have not used any data in the
manuscript. The authors have only compared the results with already openly available cosmological results.]

\section*{Conflict of Interest}

There are no conflicts of interest.

\section*{Funding Statement}

There is no funding to report for this article.

%--------------------------------------------------------------------
%	BIBLIOGRAPHY
%--------------------------------------------------------------------
%\newpage

\bibliographystyle{ieeetr} % Title is link if provided

\bibliography{bibliography}

\end{document}